\documentclass[journal=nalefd,manuscript=letter]{achemso}
\usepackage[version=3]{mhchem}

\usepackage{graphicx}
\usepackage{subfigure}
\usepackage{dcolumn}
\usepackage{bm}

\title[\texttt{achemso} demonstration]
{Manipulation and detection of spin state of Iron-Porphyrin  by dedicated chemisorption on magnetic substrates}
\author{Sumanta Bhandary}
\affiliation{Department of Physics and Astronomy, Uppsala University, Box 516,
 751\,20 Uppsala, Sweden}
 \author{Barbara Brena}
\affiliation{Department of Physics and Astronomy, Uppsala University, Box 516,
 751\,20 Uppsala, Sweden}
 \author{Pooja M. Panchmatia}
 \affiliation{School of Applied Sciences, University of Huddersfield, Queensgate,
Huddersfield. HD1 3DH, U. K.}
 \author{Iulia Brumboiu}
  \affiliation{Department of Physics and Astronomy, Uppsala University, Box 516,
 751\,20 Uppsala, Sweden}
\author{Olle Eriksson}
\affiliation{Department of Physics and Astronomy, Uppsala University, Box 516,
 751\,20 Uppsala, Sweden}
\author{Biplab Sanyal}
\email{Biplab.Sanyal@physics.uu.se}
\affiliation{Department of Physics and Astronomy, Uppsala University, Box 516,
 751\,20 Uppsala, Sweden}
 
\date{\today}
\begin{document}

%\maketitle
\begin{abstract}

%\section{Results}
{\bf One of the key factors behind the rapid evolution of molecular spintronics is the efficient realization of spin manipulation of organic molecules with a magnetic center. The spin state of such molecules may depend crucially on the interaction with the substrate on which they are adsorbed. In this letter, we demonstrate, using ab initio density functional calculations, that the stabilization of a high spin state of an iron porphyrin (FeP) molecule can be achieved via a dedicated chemisorption on magnetic substrates of different species and orientations. It is shown that the strong covalent interaction with the substrate increases Fe-N bond lengths in FeP and hence a switching to a high spin state (S=2) from a low spin state (S=1) is achieved. A ferromagnetic exchange interaction is established through a direct exchange between Fe and substrate magnetic atoms as well as through an indirect exchange via the N atoms in FeP. The mechanism of exchange interaction is further analyzed by considering structural models constructed from ab initio calculations. Finally, we illustrate the possibility of detecting a change in the molecular spin state by x-ray magnetic circular dichroism, Raman spectroscopy and spin-polarized scanning tunneling microscopy.}
\end{abstract}

Manipulation and detection of spin states of organic molecules with a magnetic centre are important issues in molecular spintronics. The change in the spin state can be achieved by external agents like temperature, light, pressure etc. Recently, it was shown \cite{sumantaprl} that a transition from S=1 to S=2 spin state of an iron porphyrin (FeP) molecule is possible by strain engineering of a defected graphene sheet. The microscopic mechanism behind this change in the spin state is identified as  the change in Fe-N bond length due to the interaction with strained graphene. The difference in the occupancy of molecular orbitals due to different bond lengths is responsible for the change in the magnetic moment of the Fe atom at the center of FeP. In this letter, we demonstrate another route to realize this effect, but now by dedicated chemisorption on a magnetic metallic substrate, which may offer an easier experimental investigation. 

Metalorganic molecules supported on magnetic substrates have attracted a lot of attention in the last few years \cite{morg}. The primary focus is to understand and manipulate magnetic exchange coupling between the 3d metal center of a metalorganic (FeP, FePc, CoPc etc.) molecule and a metallic magnetic substrate, such as Co, Ni etc. It has been shown that the exchange coupling can be tuned to be either ferromagnetic \cite{natmat07} or antiferromagnetic \cite{prl09}, depending on the chemical environment and hence on the mechanism of magnetic exchange interaction between Fe and the magnetic atom of the substrate. Sophisticated experimental tools, such as element specific x-ray magnetic circular dichroism (XMCD) \cite{natmat07} and spin-polarized scanning tunneling microscopy (SPSTM) \cite{wiesendanger} as well as accurate density functional theory calculations \cite{natmat07, prl09, wiesendanger} have played an instrumental role in uncovering the microscopic mechanisms behind these exotic phenomena.

The adsorption of metalorganics, e.g., FeP on a metallic magnetic substrate can take place in two distinct ways. One is physisorption where the molecule has a weak van der Waals interaction with the substrate without forming any chemical bond and hence, more or less preserves its gas phase properties. However, in the scenario of chemisorption, the molecule-substrate interaction  is much stronger in the presence of well established chemical bonds \cite{ehesan}. A substantial change in the geometry and electronic structure of the molecule is expected in this case due to the change in the ligand field exerted on the Fe d-orbitals. To be precise, the microscopic reason is the  enhancement of Fe-N bond lengths ( $>$ 2.04 \AA), which are larger than those in gas phase or the physisorbed molecule. Moreover, we will show below that  irrespective of surface orientation, FeP prefers to stabilize itself in such a way that it maximizes the chemical bonding between the nitrogen atoms of FeP and the underlying substrate atoms. Hence, we propose a general procedure of stabilizing  metalorganics on magnetic substrates via dedicated chemisorption to introduce a mechanical strain on the molecule which leads to a transition from a low to a high spin state. We will also show how one can detect the change in the spin states by Raman spectroscopy and SP-STM experiments.
\begin{figure}
\begin{center}
\includegraphics[scale=0.36]{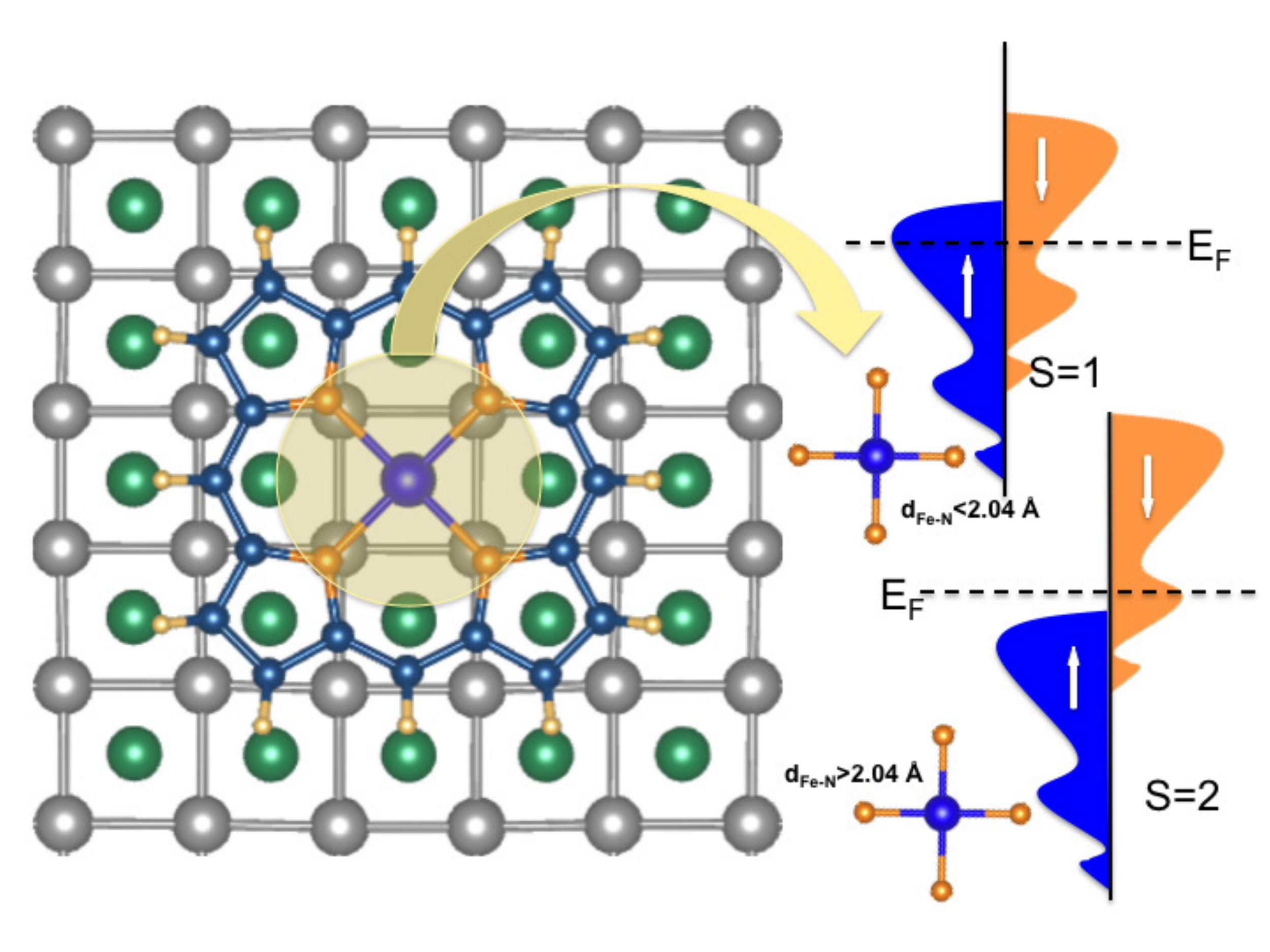}
\caption{(Color online) Geometry of a FeP molecule on a Co(001) substrate. The shaded central part demonstrates the change in the Fe-N bond-distance along with the change in the spin state. A smaller Fe-N bond length (< 2.04 \AA) leads to a smaller exchange splitting with a S=1 spin state (top right). Both the majority and minority spin channels are thus partially filled. A larger exchange splitting is observed for an increased Fe-N bond length (> 2.04 \AA), where the majority channel gets occupied completely and a high spin state (S=2) is obtained (bottom right).} 
\label{fig1}
\end{center}
\end{figure}
\par We have performed first principles density functional calculations using the VASP code \cite{kresse93prb47:R558, kresse96prb54:11169}. Plane wave projector augmented wave basis was used in the Perdew Becke Ernzerhof generalized gradient approximation (PBE-GGA) for the exchange correlation potential. The plane wave cut off energy used was 400 eV. A 3x3x1 Monkhorst Pack k-point set was used for the integration in the Brillouin zone. Atoms were relaxed until the Hellman Feynman forces were minimized up to 0.01 eV/\AA. We have used 7x7 lateral supercells of the magnetic substrate for the (001) and (111) orientations and a 6x7 supercell was used for the (110) surfaces. In all cases, a slab of three layers and 21 \AA~of vacuum length perpendicular to the substrate were considered. For geometry optimizations, the atoms in the lowest layer were kept fixed and all the other atoms in the slab as well as in the FeP molecule were allowed to relax. To account for the electronic correlations in the narrow d-states of Fe in the FeP molecule, we have used GGA+U approach where a Coulomb interaction term is added according to the mean field Hubbard U formalism \cite{anisimov}. The values of the Coulomb parameter U and the exchange parameter J were chosen to be 4 eV and 1 eV respectively as these values correctly reproduce the electronic structure and magnetism of FeP in the gas phase \cite{chemphys}. For all cases, we have used a semi-empirical approach \cite{grimme} to account for van der Waals interaction between FeP and the substrates.
\par The generality of the change in molecular spin state has been studied by having different types of substrates and their orientations, i.e., (100), (110) and (111). Magnetic substrates such as Co and Ni with partially filled d-orbitals are chosen to enhance the probability of chemisorption and also to allow for a study of the mechanism of magnetic exchange interactions. Different surface orientations are chosen to provide different crystallographic environments to the molecule. Ni surfaces with hexagonal (111), rectangular (110) and square (100) facets are considered along with a Co (100) surface. The (100) surface orientation of Co and Ni provides the most symmetric substrate (square) for the molecule whereas the rectangular (110) or hexagonal (111) surface provides a less symmetric environment. This  indeed affects the stabilization as well as the geometry of the chemisorbed FeP molecule. To be more specific, on a square substrate, Co/Ni atoms interact with all four N atoms of FeP in a similar fashion whereas on a rectangular surface, the chemical interaction along two perpendicular directions are different. This is also true for the hexagonal surface.  We have also considered a configuration where FeP is rotated by 45$^\circ$ with respect to the substrate in order to have a less direct interaction between FeP (predominantly with ligand N) and the first substrate layer. Even though we obtain a high spin state for all chemisorbed situations, chemical stabilization depends on the  rotation of the molecule with respect to the substrate, as will be discussed below. It is indeed observed that the maximization of the molecule-substrate direct chemical interaction increases the  stability of adsorption. In all studied systems, can one observe that the interaction between N states of the FeP molecule and the surface orbitals of the substrate is maximized in the ground state configuration.

\begin{table}[!hbp]
\begin{center}
\caption{\label{tab1} Calculated exchange energies ($E^{ex}=E^{AFM}-E^{FM}$)  and relative energies w.r.t the ground state (in parenthesis) in eV are shown for different substrates and their orientations. Note that the positive $E^{ex}$ indicates that the FM alignment is stable. The configurations denoted by {\sl TOP}, {\sl TOP-R}, {\sl HOLLOW} and {\sl HOLLOW-R} are defined in the text. For the Ni(111) surface, the {\sl TOP} configuration is the only relevant one.}
\begin{tabular}{|c|c|c|c|c|} \hline \hline
Surface &  {\sl TOP} & {\sl TOP-R} & {\sl HOLLOW} & {\sl HOLLOW-R} \\ 
\hline 
Ni (110) & 0.04 (2.5) & 0.09 (0.0) & 0.43 (1.06) & 0.13 (1.88) \\
Ni (111) & 0.13 & -- & -- & -- \\
Ni (001) & 0.07 (2.17) & 0.2 (3.58) & 0.16 (0.0) & 0.10 (3.13) \\
Co (001) & 0.39 (2.49) & 0.49 (3.6) & 0.75 (0.0) & 0.32 (4.38) \\
\hline \hline
\end{tabular}
\end{center} 
\end{table}

%\begin{table}[!hbp]
%\begin{center}
%\caption{\label{tab1} Binding energies (in eV) of FeP chemisorbed on different surfaces are shown. In parentheses,  the calculated exchange energies ($E^{ex}=E^{FM}-E^{AFM}$)  in eV are also given.}
%\begin{tabular}{|c|c|c|c|c|} \hline \hline
%Surface &  Top & Top-rotated & Hollow & Hollow-rotated \\ 
%\hline 
%Ni (110) & -2.973 (0.04) & -5.468 (0.09) & -4.409 (0.43) & -3.582 (0.13)  \\
%Ni (111) & -2.608 (0.13) & & &  \\
%Ni (001) & (0.07) & (0.18) & (0.16) & (0.10)  \\
%Co (001) & (0.39) & (0.49) & (0.69) &  \\
%\hline \hline
%\end{tabular}
%\end{center} 
%\end{table}

\par The chemisorbed FeP is found to undergo  a mechanical strain originating from the molecule-surface interaction, primarily by the  N-Ni/N-Co direct chemical bonding, which leads to a stretching of the Fe-N bond lengths of FeP. The stretching can be symmetric (e.g., for a  (001) surface) or asymmetric (e.g., for a (110) surface) along different in-plane crystallographic directions depending on the orientation of the surface underneath. Nevertheless, all the stretched geometries yield  a high spin state (S=2) of FeP by having the individual Fe-N bond length exceeding 2.04 \AA. This structural change in FeP has a direct impact on its electronic properties.
The four N atoms bonded to the central Fe atom (in a Fe$^{2+}$ state) provides a square planar crystal field. The p-d hybridization between N-2p and Fe-3d orbitals shifts  the $d_{x^2-y^2}$ level higher up in energy due to strong $\sigma$-type hybridization, whereas other orbitals feel a weak $\pi$-type hybridization and stay lower in energy. Along with this covalent contribution, a point charge contribution shifts the $d_{x^2-y^2}$ level up due to a strong Coulomb repulsion. Both these parts of the crystal field (CF) are strong when the Fe-N bond length is smaller, e.g. in the gas phase. Hence, the CF dominates over the intra atomic exchange and a low spin S=1 state is observed with 4 spin-up and 2 spin-down electrons. Now as the Fe-N bond length is stretched, the CF strength decreases resulting in the lowering of the $d_{x^2-y^2}$ orbital (\ref{fig1}). The electrons now occupy Fe-3d orbitals following Hund's first rule by filling up all majority spin levels followed by one level in the minority spin channel. Even though we have observed a strong bonding between Fe $d_{z^2}$   and a surface atom, as the spin state change is mostly governed by the shift of the $d_{x^2-y^2}$ orbital, CF effects explain the spin state of the chemisorbed FeP molecule.

%The reason being the occurrence of N-2p orbitals quite low in energy compared to 3d levels of Fe. In this scenario, the bonding and anti bonding levels due to the direct overlap, will contain dominantly N-2p and Fe-3d respectively. Now, the bonding level will appear far low in energy whereas the anti bonding sates will appear close to the Fermi-level, being the most crucial ones in this regard. So, ordering of the Fe-d levels as well as the exchange splitting will primarily be governed by the ligand field (LF) provided by N-atoms, even though a strong hybridization is observed when Fe atom sits on top of a surface atom. Irrespective of the relative position of d-orbitals, as shown in (Fig.~1), when the Fe-N bond length  is stretched, the exchange splitting is increased due to a downward shift of $d_{x^2-y^2}$ orbital in energy. This results in the  complete occupation of spin-up channel and a partial occupation of spin-down channel. On the contrary, both spin channels were partially occupied for the situation with a low spin state (S=1).

\begin{figure}[!hbp]
\begin{center}
\includegraphics[scale=0.36]{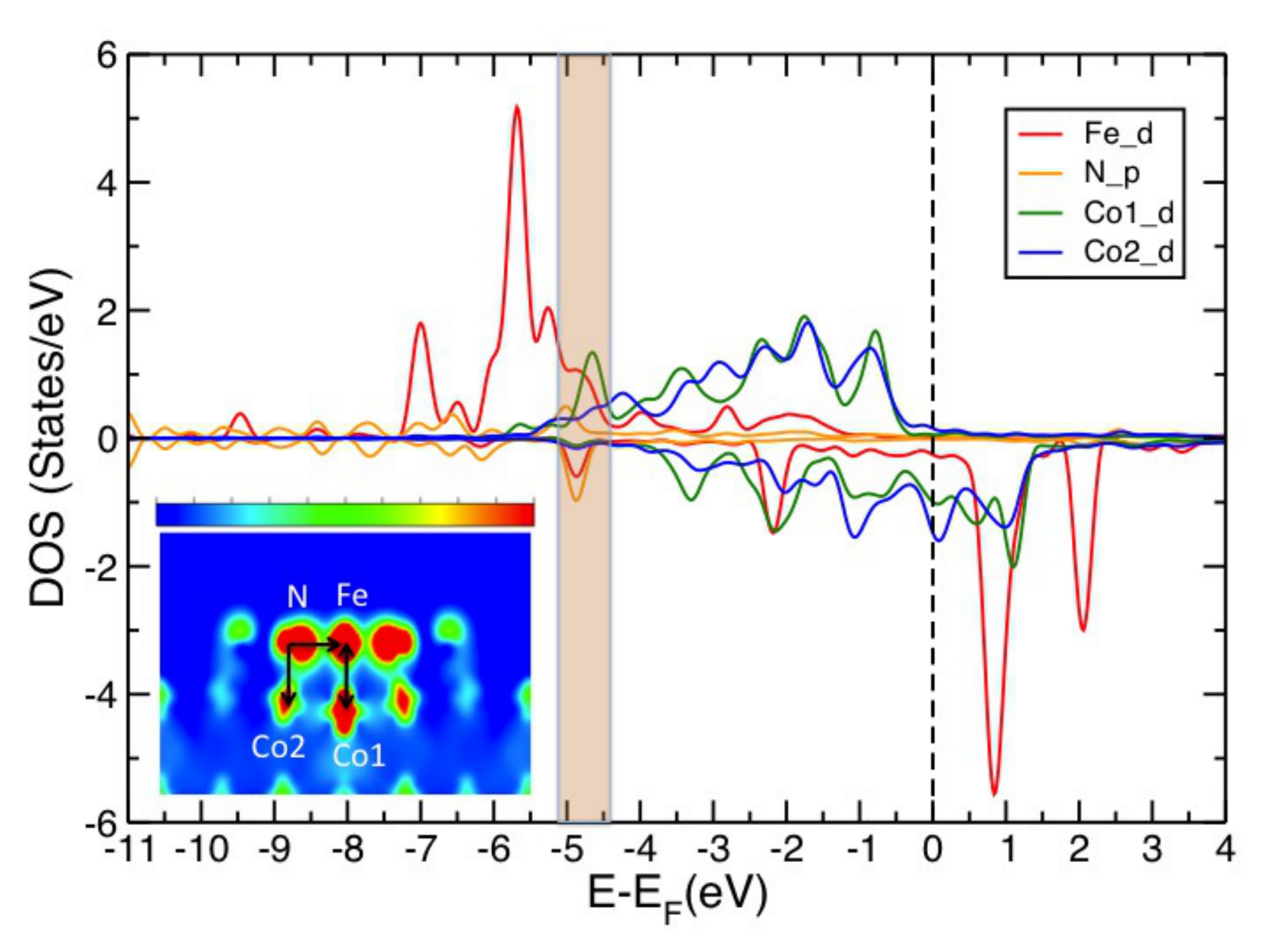}\\
(a)\\~\\
\includegraphics[scale=0.20]{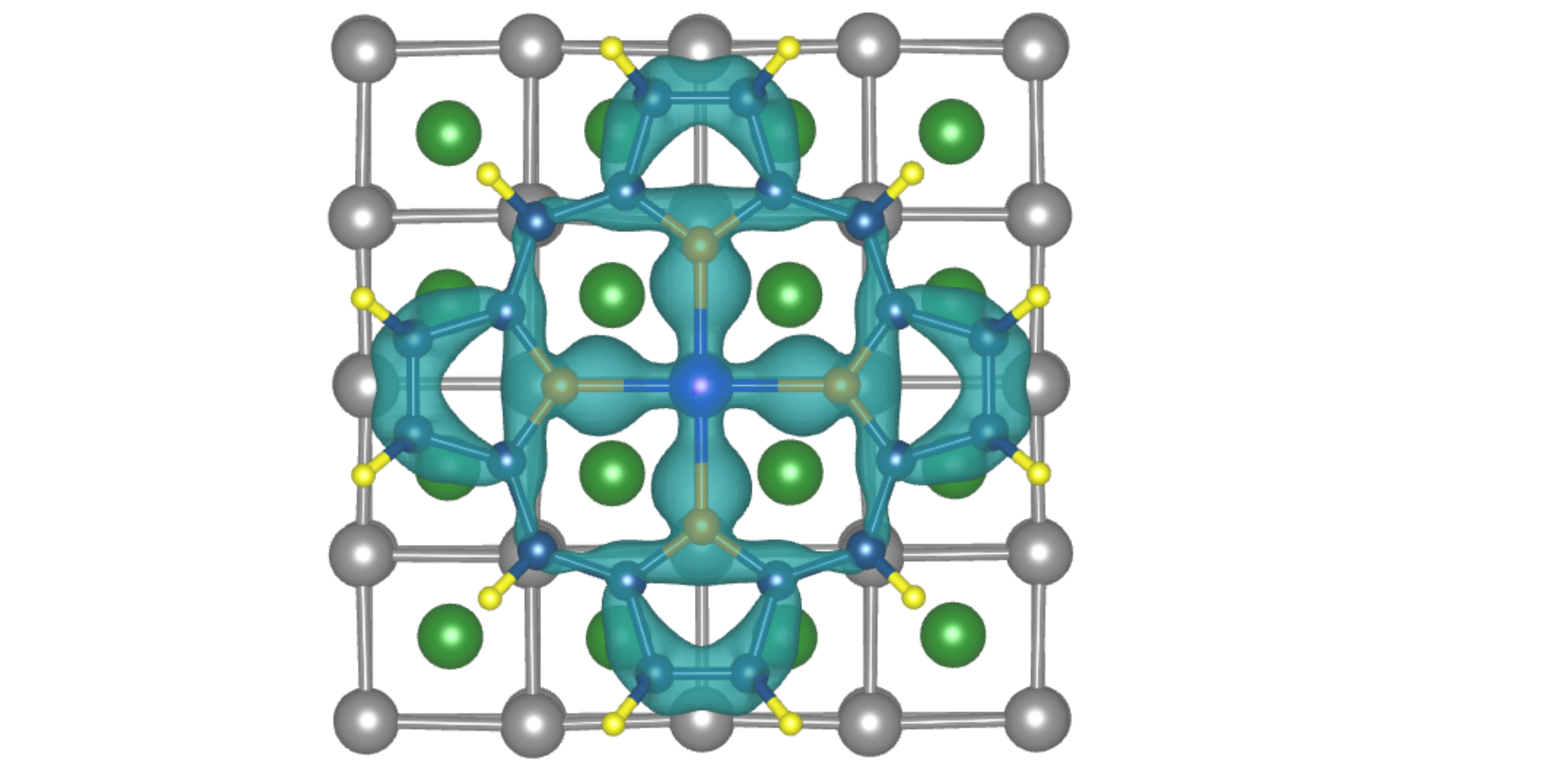} 
 \includegraphics[scale=0.18]{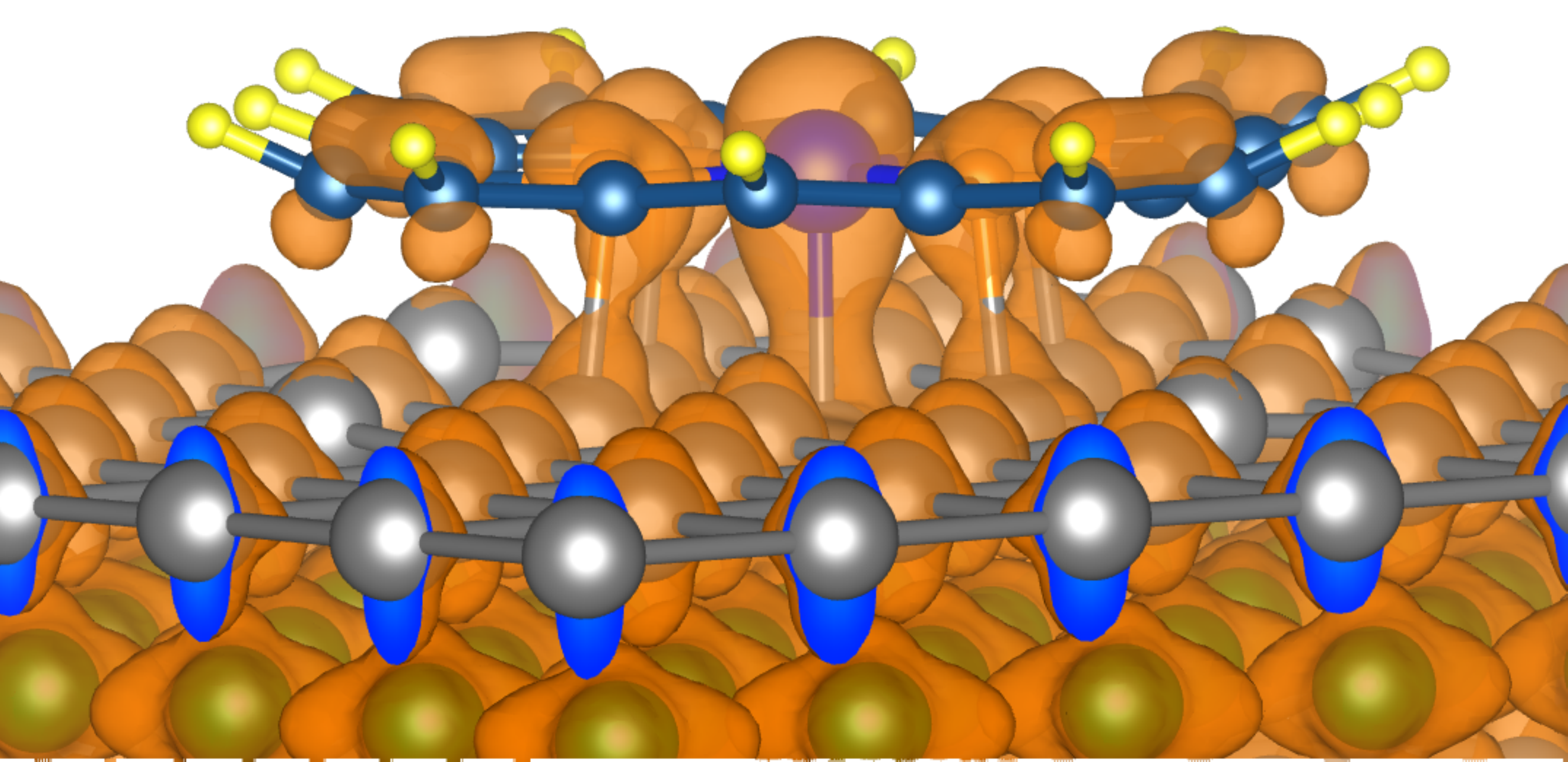}\\
 (b) \hskip 5cm (c) \\~\\
\caption{(Color online) (a) Spin-polarized atom and orbital-resolved DOS are shown for FeP on Co (001) in the {\sl TOP} position. In the inset, a charge density cut in the (001) plane is shown for an energy window indicated as the shaded region in the DOS plot. Co1 and Co2 are the atoms of the Co substrate, situated just below Fe and N atoms of FeP respectively. (b) Top view of the magnetization density isosurface shown for the same energy interval to show the in-plane orbitals. (c) Side view of the magnetization density to show the out-of-plane orbitals taking part in the exchange interaction between FeP and the Co substrate.} 
\label{fig2}
\end{center}
\end{figure}
\par Due to chemisorption, the Fe moment has a strong exchange interaction with the substrate Co/Ni moments. The magnetic exchange coupling energies ($E^{ex}$=E$^{AFM}$-E$^{FM}$) obtained in the ground state structures on each surface are given in \ref{tab1}. $FM$ and $AFM$ denote ferromagnetic and antiferromagnetic alignments between the Fe and substrate moments respectively. It should be noted that both direct and indirect superexchange interactions give rise to a ferromagnetic exchange coupling in these systems. When both Fe and N atoms are on top of surface atoms, a (Co/Ni)-Fe direct exchange as well as (Co/Ni)-N-Fe indirect superexchange interactions are operational. This situation is denoted as the {\sl TOP} configuration, for which spin-polarized DOS, energy-resolved charge and magnetization densities are shown in \ref{fig2}. The charge density is plotted (inset of \ref{fig2}(a)) in an energy window (shaded part in \ref{fig2}(a)) of 0.23 eV where the hybridization between Fe-N, N-Co and Fe-Co orbitals are dominant.  The cross section of the total charge density in the (100) plane (\ref{fig2}(a)) indicates a substantial overlap between the orbitals mentioned above. The magnetization densities shown in \ref{fig2}(b) and \ref{fig2}(c) indicate a similar overlap between in-plane and out-of-plane orbitals carrying magnetic moments. To be precise, the $d_{x^2-y^2}$ orbital of Fe hybridizes with the orbitals of N in the plane of FeP as seen in the spin-down channel (\ref{fig2}(b)). The out of plane hybridization between Fe-Co and N-Co via $d_{z^2}$ and $p_z$ orbitals are observed in the spin-up channel (\ref{fig2}(c)). Therefore, the charge and magnetization densities show the signatures of a direct exchange as well as an indirect superexchange interaction between Fe and substrate Co atoms. 

Besides the {\sl TOP} configuration mentioned above, FeP can be adsorbed in other ways, e.g., {\sl TOP-R}, {\sl HOLLOW} and {\sl HOLLOW-R} configurations. For the nomenclatures, the readers are referred to the Supplementary Information. 
If FeP is rotated by 45$^{\circ}$ angle with respect to the substrate (denoted as {\sl TOP-R}/{\sl HOLLOW-R}), N-atoms are no longer placed on top of substrate atoms. In this situation the effect of indirect superexchange is minimized and the direct exchange between the Fe atom in FeP and the substrate atoms dominates. But the absence of N-substrate bonds pushes the molecule further apart from the substrate which in turn, has an  effect on  the direct exchange. On the Co(100) surface, FM exchange is favored over AFM by 0.39 eV for the {\sl TOP} configuration and by 0.49 eV for {\sl TOP-R} configuration. For the {\sl HOLLOW} configuration, the exchange coupling energy is 0.69 eV favoring a FM alignment. Therefore, irrespective of chemical species and orientation, all surfaces show FM coupling with FeP. Apart from the exchange energies, the relative stabilities are also indicated in \ref{tab1} through the total energies w.r.t. the ground state energy for each surface type and orientation. It is clearly observed that the most stable configuration is the {\sl HOLLOW} position for both Co and Ni (001) surfaces whereas the {\sl TOP-R} configuration is the most stable for a Ni(110) surface.  As mentioned before, stability is primarily dictated by the chemical bonds between the atoms in the molecule and in the first surface layer. Also, a difference in behavior in the relative stabilities between Co and Ni surface is observed due to the difference in number of $d$-electrons and hence the strength of bonding.%[{\bf may be we do not need this line ::  One thing should be noted that on Ni(110) surface, FeP moves laterally to maximize the Ni-N bond. This results to a deformation relative to the ideal rotated situation}]. 
\begin{figure}[!hbp]
\begin{center}
\includegraphics[scale=0.4]{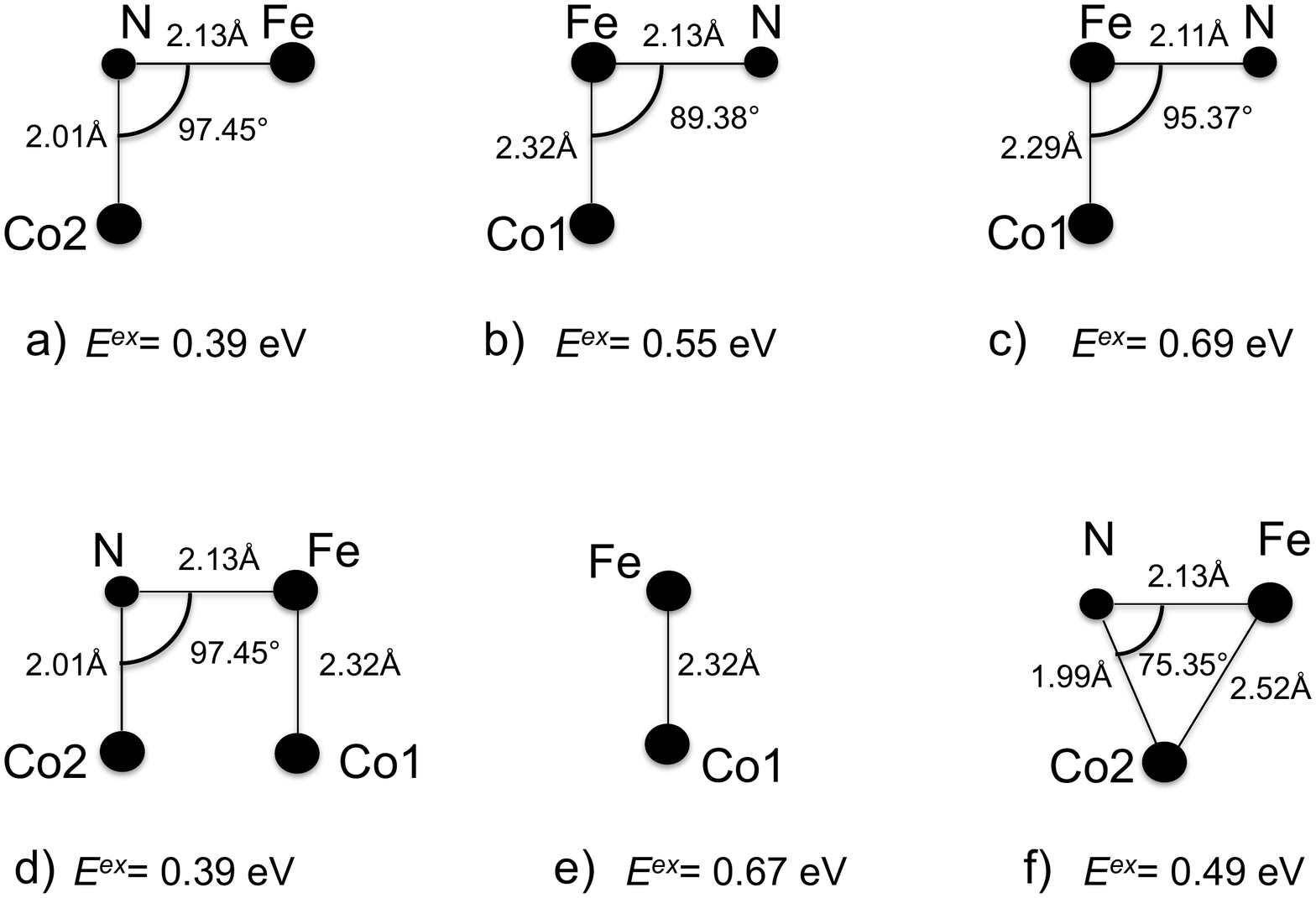}
\caption{Structural models extracted from fully optimized DFT calculations of FeP on different substrates indicated in the text. The geometries were not optimized further for these models those are used to calculate the total energies for FM and AFM alignments between Fe and Co moments. The energy differences between FM and AFM alignments are shown for different models considered.} 
\label{fig3}
\end{center}
\end{figure}
\par In order to have a better understanding of the exchange mechanism, we have performed several total energy calculations with the structural models extracted from the optimized geometries in our full DFT calculations. We have chosen the relevant parts of the optimized geometries to build the structural models, which are shown in \ref{fig3} for the {\sl TOP} configuration. As we have already stated, there are two kinds of possible surface to molecule exchange interactions, viz., (Co/Ni)-Fe direct exchange and (Co/Ni)-N-Fe indirect superexchange. \ref{fig3}(a) represents the indirect superexchange between Fe and Co  via N whereas \ref{fig3}(b) represents the Fe-Co direct exchange. Here Co1(Co2) is the Co atom just below Fe (N). As mentioned earlier, a rotation by 45$^{\circ}$ increases the distance between Fe and Co.  \ref{fig3}(c) mimics the situation having a direct exchange but with larger Fe-Co distance. \ref{fig3}(d) represents a model, that includes both direct and indirect exchange while \ref{fig3}(e) represents Fe-Co direct exchange in the absence of N. \ref{fig3}(f) is taken from the optimized {\sl HOLLOW} configuration. In this case, the Fe-N-Co angle is not close to 90$^{\circ}$ but 75.35$^{\circ}$ due to the geometry of the surface underneath. Based on our calculations, we can conclude that both direct and indirect exchange interactions are important to consider with varying importance based on particular situations of chemisorption. %as N-Fe-N distance is now to be compared with Co-Co next nearest neighbor distance unlike 2x(Co-Co nearest neighbor distance) for Top-situation. As the distances matches quite well on top-situation the Fe-N-Co angle becomes 97.45$^{\circ}$. As we can see, both the calculations predict FM interaction and is favored by both  direct exchange and indirect  superexchange. Also, as the Fe-Co direct distance increases, the strength of FM exchange increase. This observation agrees with the calculations with the full geometry. }

 %Total energy calculations have been done for FM and AFM alignments between Fe and Co atoms (flipping Fe-moment) and the results are presented in \ref{fig3}.
%The exchange mechanisms are different when the Fe atom of FeP is in a {\it TOP} or a {\it HOLLOW} position. Fe-N-Co/Ni angle is 97.45$^{\circ}$ for {\it TOP}  whereas it changes to 75.34$^{\circ}$ at the {\it HOLLOW} position. Nevertheless all the direct or indirect exchange mechanisms favor a FM exchange. Also, the model calculations show that as Fe-Co1 distance increases, the strength of FM exchange increase. This observation agrees with the calculations on the full geometry.  
\par\begin{figure}
\begin{center}
\includegraphics[scale=0.6]{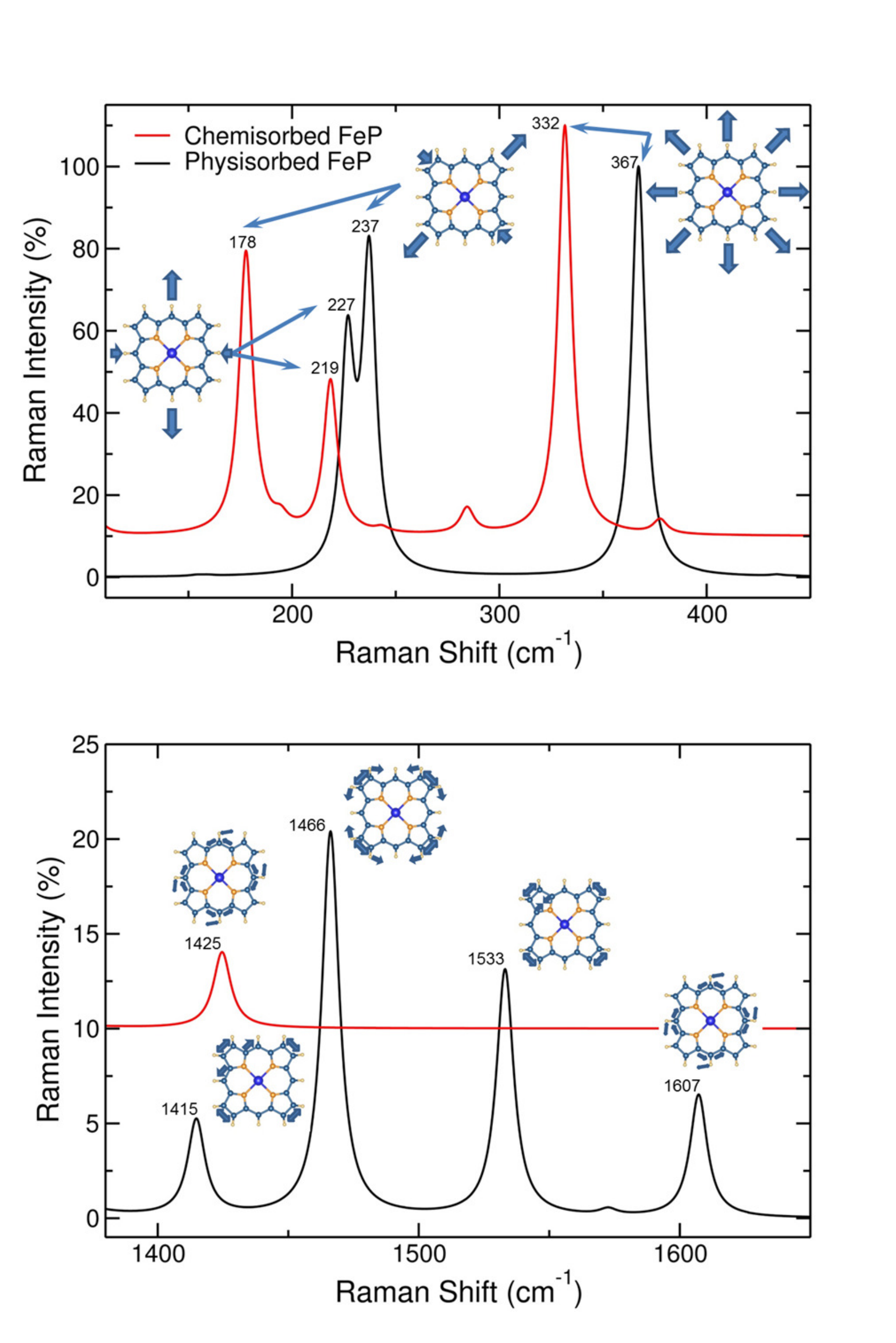}
\caption{(Color online) Calculated Raman intensities for FeP in chemisorbed and physisorbed geometries. In the upper panel, Raman intensities are shown for shifts in the range 110-450 cm$^{-1}$ whereas the lower panel contains wave numbers in the range 1380-1650 cm$^{-1}$. In both panels, the vibrational modes corresponding to the peaks are denoted along with the corresponding wave numbers.} 
\label{fig4}
\end{center}
\end{figure}
 \par Although the spin moment remains similar in all chemisorbed situations, a deformation in the spin density is expected to be quite high for a low symmetry structure, leading to a large value of spin dipole moment $\langle T_{z} \rangle$ \cite{sumantaprl}, where  $\langle T_{z} \rangle$ is the expectation value of the z-component of spin-dipole operator $T$. Moreover spin densities differ due to different ligand fields exerted on FeP by different surfaces and their orientations. The discussion on spin dipole moment is quite relevant for the XMCD measurements where the measured effective moment contains both spin and spin dipole moments. We have calculated the spin dipole moments by following the method suggested by Freeman {\it et al.} \cite{freeman}. The calculated values of 7$\langle T_{z} \rangle$ are 0.82 $\mu_{B}$, 1.08  $\mu_{B}$ and 0.79 $\mu_{B}$ for FeP on Co(001), Ni(111) and Ni(110) surfaces respectively. It is clear from the signs and values of $\langle T_{z} \rangle$ that the effective spin moment defined as $S_{eff}=2 \langle s_{z} \rangle + 7 \langle T_{z} \rangle$, can be highly affected by $\langle T_{z} \rangle$. This change in the value of $S_{eff}$ should be detectable in XMCD measurements.
 
% \section {Detection}
Besides the spin dipole moments measured in XMCD experiments, we propose two additional experimental techniques to detect the change in the spin state of FeP, (i) Raman spectroscopy and (ii) spin-polarized STM (SPSTM). We shall first discuss our results for Raman intensities. For these calculations, we have considered the optimized geometries of FeP in both chemisorbed and physisorbed situations in presence of the substrates and have calculated the vibrational normal modes of FeP in those optimized geometries only in the gas phase. The Gaussian 09 program \cite{g09} was used at the B3LYP level of theory with the 6-31+G(d,p) basis set. In these calculations, the geometries were taken from the optimized structures of FeP on Co(001) for chemisorbed and physisorbed situations. The calculated wave numbers above 1000 cm$^{-1}$ were scaled using the correction factor 0.9614 \cite{corr} . Raman Intensities were calculated using the expression \cite{raman}: $$I (\nu_{i})=\frac{fS_{i}(\nu_{0}-\nu_{i})^{4}}{\nu_{i}(1-e^{\frac{-hc\nu_{i}}{k_{B}T}})},$$ where $f$ is a scaling factor (1.0 in this case), $S_{i}$ are the calculated Raman activities, $\nu_{0}$ is the wave number corresponding to the excitation laser line in cm$^{-1}$ (we used 457.9 nm as it is in the range used to measure the resonance Raman spectrum of porphyrins \cite{fepraman}), $\nu_{i}$ is the calculated vibrational frequency, $h$, $c$ and $k_{B}$ are Planck's constant, speed of light and Boltzmann's constant respectively, while $T$ is the room temperature (293 K).
\begin{figure}
\begin{center}
\includegraphics[scale=0.4]{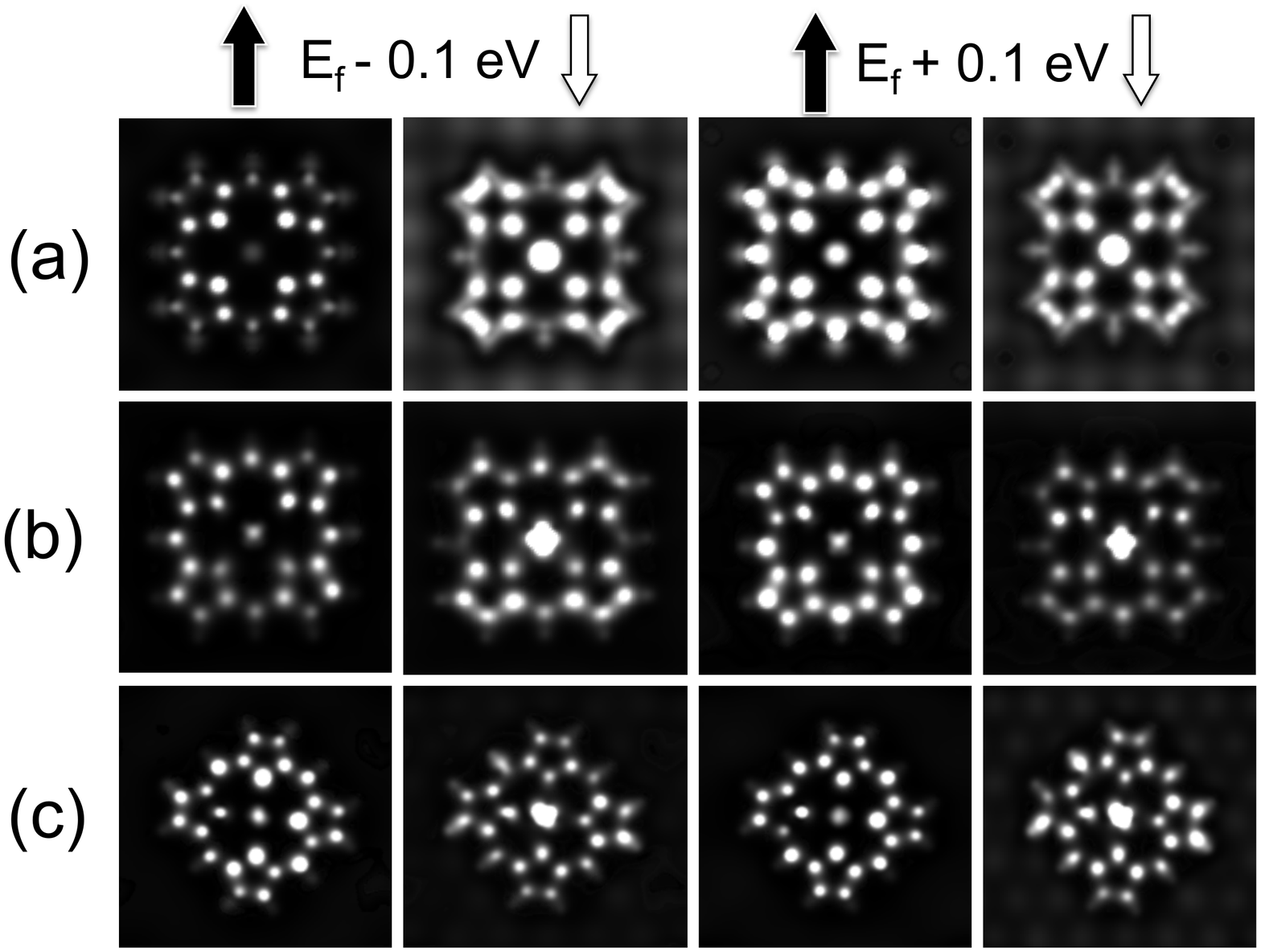}
\caption{Calculated spin-polarized STM images are shown for bias voltages +/- 0.1 V for FeP chemisorbed on (a) Co(001), (b) Ni(110) and (c) Ni(111) surfaces. STM images for spin-up and spin-down channels are shown separately, indicated by up and down arrows respectively. E$_{\mathsf{f}}$ denotes the Fermi energy.} 
\label{fig5}
\end{center}
\end{figure}
\par The Raman spectra of the triplet (S=1) and quintuplet (S=2) spin states differ substantially both in the peak intensity and peak position, as evident from \ref{fig4}. There are two spectral windows where these differences are significantly observed. First of all, the most intense peaks of the FeP (both in the case of the chemisorbed and physisorbed molecule) appear in the range of 110 - 450 cm$^{-1}$, shown in the upper panel of \ref{fig4}. The highest band in this spectral window corresponds to a breathing vibration of the entire molecule and it is shifted to lower wave numbers in the case of the chemisorbed FeP as compared to the physisorbed one by approximately 35 cm$^{-1}$. A shift of this magnitude should easily be observed in Raman experiments \cite{aroca}. A shift of even larger magnitude occurs in the band at 237 cm$^{-1}$ of the physisorbed FeP. It is assigned to a symmetric stretching vibration of the entire molecule and it appears 60 cm$^{-1}$ lower in the chemisorbed situation. The third major difference in this spectral window is the inversion of the two lower Raman peaks. The less intense peaks correspond to the two spectra representing diagonal symmetric stretching vibrations of the entire molecule. In the range of 1380-1650 cm$^{-1}$ (shown in the lower panel of \ref{fig4}), three peaks occur with relatively high intensity in the physisorbed FeP that have correspondence in the chemisorbed molecule at much lower wave numbers. For instance, the physisorbed FeP band at 1607 cm$^{-1}$ corresponds to the same C-C-C asymmetric stretching coupled with C-H bending that is assigned to the 1425 cm$^{-1}$ chemisorbed FeP band. The other two peaks are assigned to in-plane C-C stretching vibrations and C-N-C bendings. Thus, this spectral window would be ideal to experimentally detect the switching between physisorbed and chemisorbed states, as the disappearance/appearance of these three bands could be easily monitored.
\par The other detection method we would like to propose is the SPSTM technique. We have simulated STM images using Tersoff-Hamman \cite{stm} approach for spin-up and spin-down channels separately. Also, both filled and empty state images are shown for a bias voltage of 0.1 V. A similar method has been used by Brede {\it et al.} \cite{wiesendanger} to simulate SPSTM images for CoPc adsorbed on a Fe substrate. From \ref{fig5}, it is clear that for this value of bias voltage, only the molecular states are observed and no signature of surface states is seen. Also, one clearly observes distinct differences in the features of the images for spin-up and spin-down channels. For example, in the Co(001) case, the spin-down channel shows a bright intensity in the region of the Fe center, whereas it is absent in the spin-down channel for the filled state images. This is true for Ni(110) surface as well. All these changes observed in the SP-STM images should be useful for detecting the spin state.
\par In summary, we have demonstrated by ab initio density functional calculations that a dedicated chemisorption of an FeP molecule on magnetic substrates of different species and orientations can switch the spin state of the molecule from S=1 to S=2. This change is brought by an increase in the Fe-N bond lengths in FeP, induced by the strong covalent interactions between the chemisorbed molecule and the substrate atoms. The ground state geometries correspond to the maximization of the chemical bonds between N atoms of FeP and the underlying substrate atoms. Moreover, we have analyzed the mechanisms of exchange interaction by simple structural models constructed from our DFT calculations and demonstrated that both superexchange and direct exchange are operational in mediating a ferromagnetic interaction between Fe and the substrate magnetic atoms. Finally, we show that it is possible to detect the change in the spin state of FeP due to chemisorption by XMCD, Raman spectroscopy and spin-polarized scanning tunneling microscopy. 

%\section*{Method}

\section*{Acknowledgements}
We gratefully acknowledge financial support from the Swedish
Research Council (VR). O.E. is in addition grateful to the KAW foundation and the ERC(project 247062 - ASD) for support. P.P would like to acknowledge membership of the UK's HPC Materials Chemistry Consortium, which is funded by EPSRC (Grant No. EP/F067496). We also acknowledge SNIC-UPPMAX, SNIC-HPC2N and SNIC-NSC centers under the Swedish National Infrastructure for Computing (SNIC) resources for the allocation of time in high performance supercomputers. 

\section{Supplementary Information}
In the Supplementary Information,  we provide detailed structural data obtained from our calculations. As stated in the main article, stability, spin state and magnetic interaction of FeP with the surfaces depend largely on the surface orientation as well as the orientation of the molecule itself with respect to the surface underneath. To give an example, the (001) surfaces of Co and Ni provide a symmetric crystallographic coordination to FeP, contrary to (110) and (111) surfaces. To demonstrate the effects of orientations and resulting bonding situations, we have chosen the case of Co(001) surface. We will demonstrate four possible configurations on top of a (001) surface, viz. (i) {\sl TOP}, (ii) {\sl TOP-R}, (iii) {\sl HOLLOW} and (iv) {\sl HOLLOW-R}. Also, in the following figures, chemical bonds less than 2.5\AA ~ are shown only to make the main content clear.
 \par
\begin{figure}[!hbp]
\begin{center}
\includegraphics[scale=0.32]{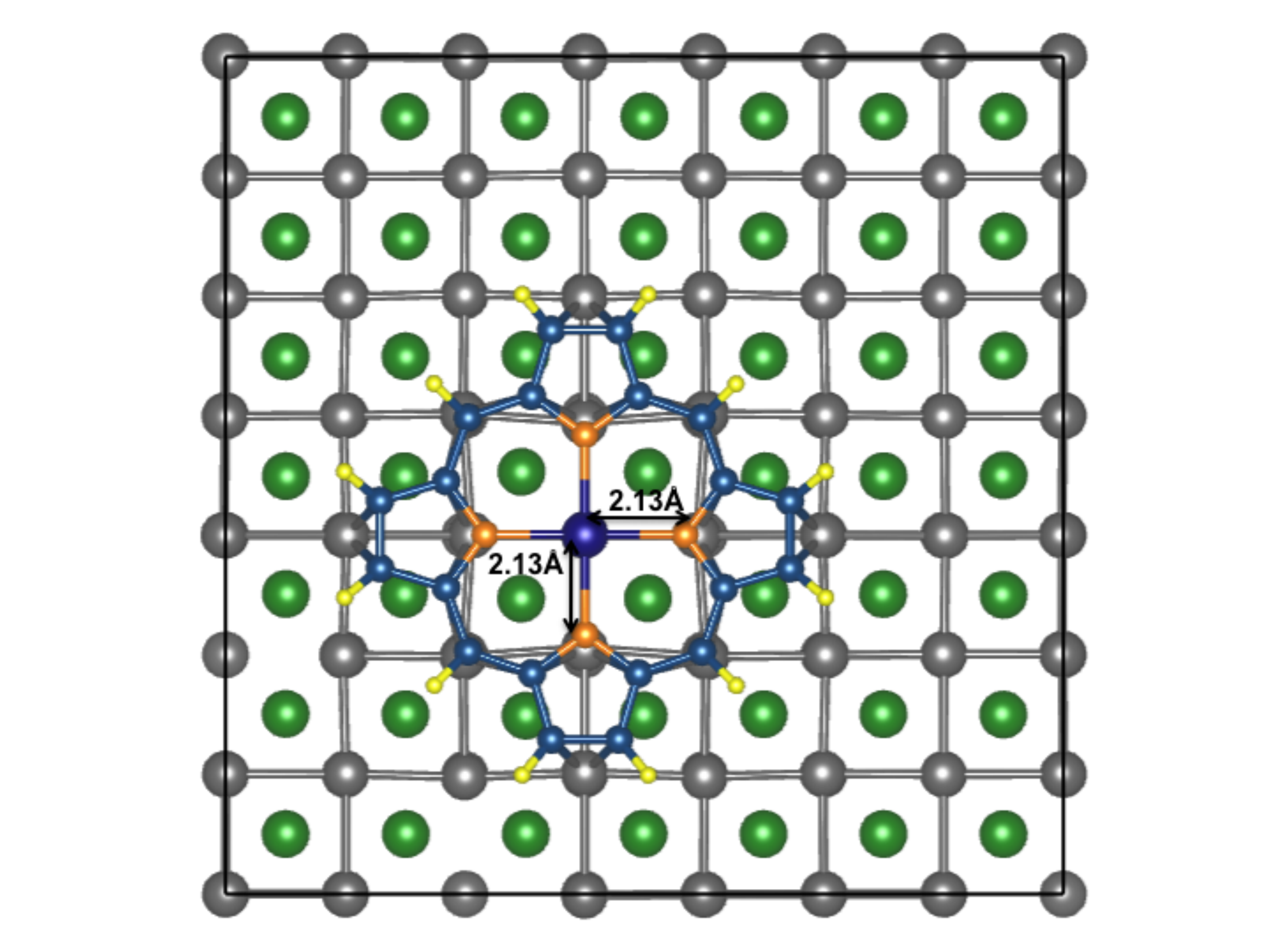} 
 \includegraphics[scale=0.32]{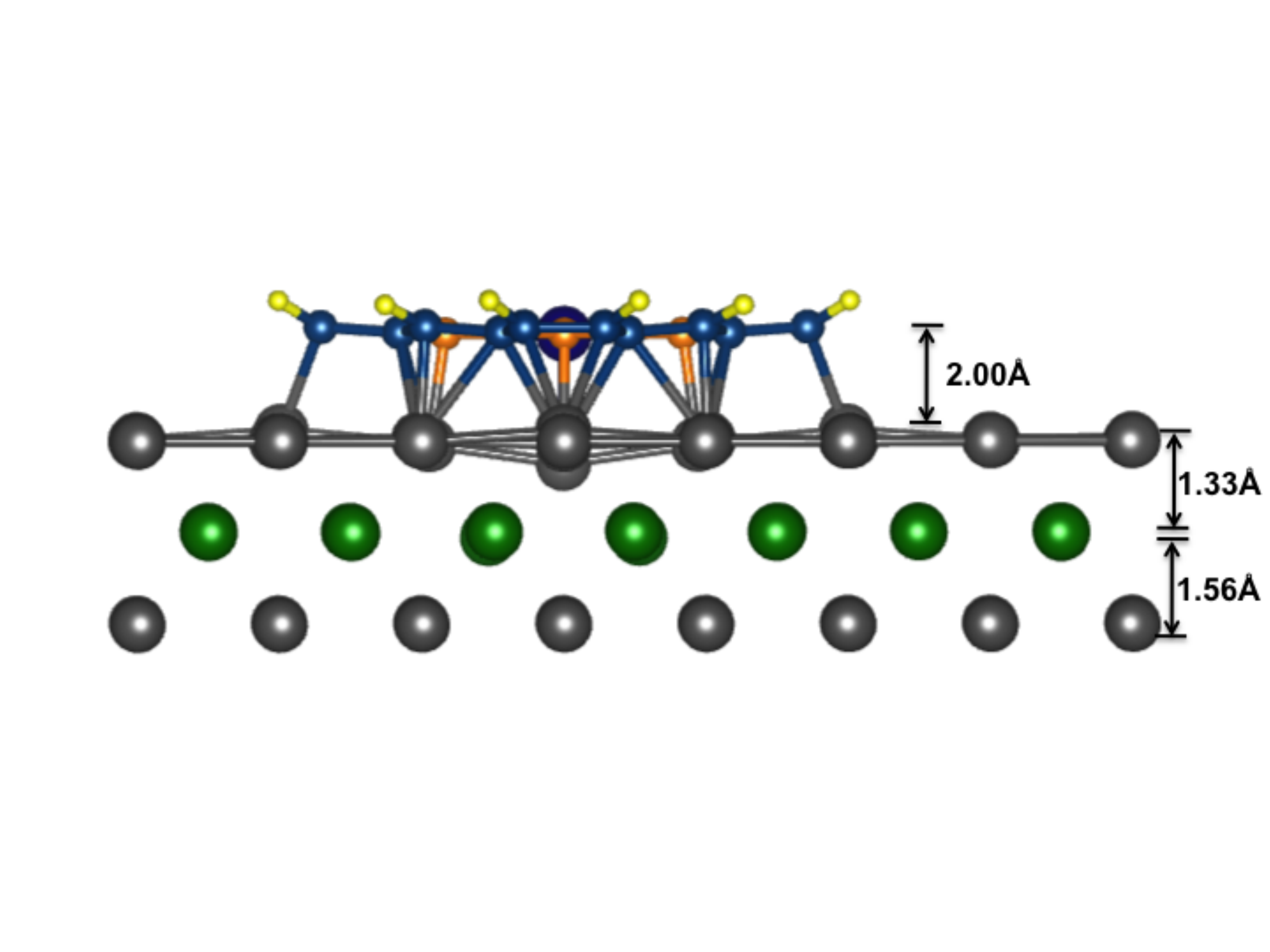}\\
 (a) \hskip 8cm (b) \\~\\
\caption{(Color online) FeP molecule on Co (001), {\sl TOP} position. The gray and green balls represent Co substrate atoms. Two different colors are used to  present different layers clearly. C atoms are shown in steal, while Fe, N and H atoms are represented with blue, orange and yellow colors respectively. (a) Top view, where Fe-N bond lengths are shown. (b) Side view, where minimum distances between different layers are shown.} 
\label{figsi1}
\end{center}      
\end{figure}

  (i) {\sl TOP}: FeP can be chemisorbed in such a way that the Fe center sits exactly on top of a surface Co atom making a direct bond between them. The four N's bonded to Fe also orient themselves so that each of them makes direct chemical bonds with one Co atom on the surface. We denote this particular geometry as the {\sl TOP}-configuration (\ref{figsi1}).  The Fe-N bond length  in the FeP-gas phase molecule is 2.0 \AA~where a Co-Co nearest neighbor distance is 2.5 \AA. To form a direct Fe-Co bond, as well as a N-Co bond, which helps in stability, the molecule has to stretch. Therefore, the Fe-N bond length is stretched up to 2.13 \AA~(\ref{figsi1}) when adsorbed on the (001) surface. 

Moreover, a distortion is observed mostly in the surface layer of Co where Co-Co bond length is contracted to 2.44 \AA. Thus mutual stretching and contraction occur during the formation of the chemical bonds. As the Fe-N and Co-Co bonds are not exactly the same, an Fe-N-Co bond angle larger than 90$^{\circ}$ is established, 97.44 to 98.16$^{\circ}$, to be precise. Fe-Co direct bonding occurs at 2.32 \AA ~even though the minimum distance of the molecule from the surface layer is 2.0 \AA, as shown in \ref{figsi1}(b). The Co atom, just below Fe is pushed down and this also decreases the minimum distance between surface Co-layer and the second one down to 1.33 \AA~ while the minimum distance between 2$^{nd}$ and 3$^{rd}$ Co layers remains much higher, 1.56 \AA~ (\ref{figsi1}1(b)).

\begin{figure}[!hbp]
\begin{center}
\includegraphics[scale=0.32]{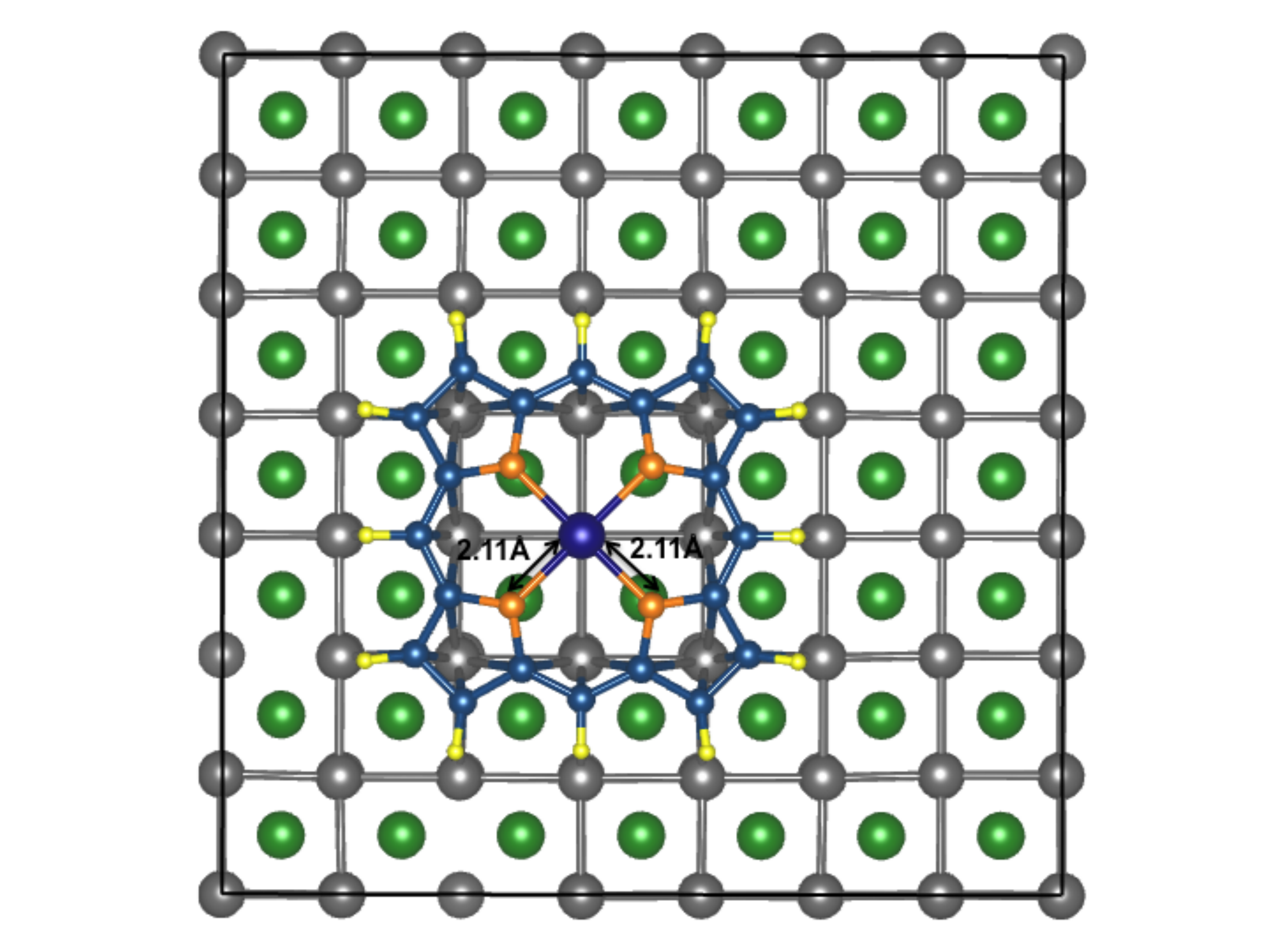} 
 \includegraphics[scale=0.32]{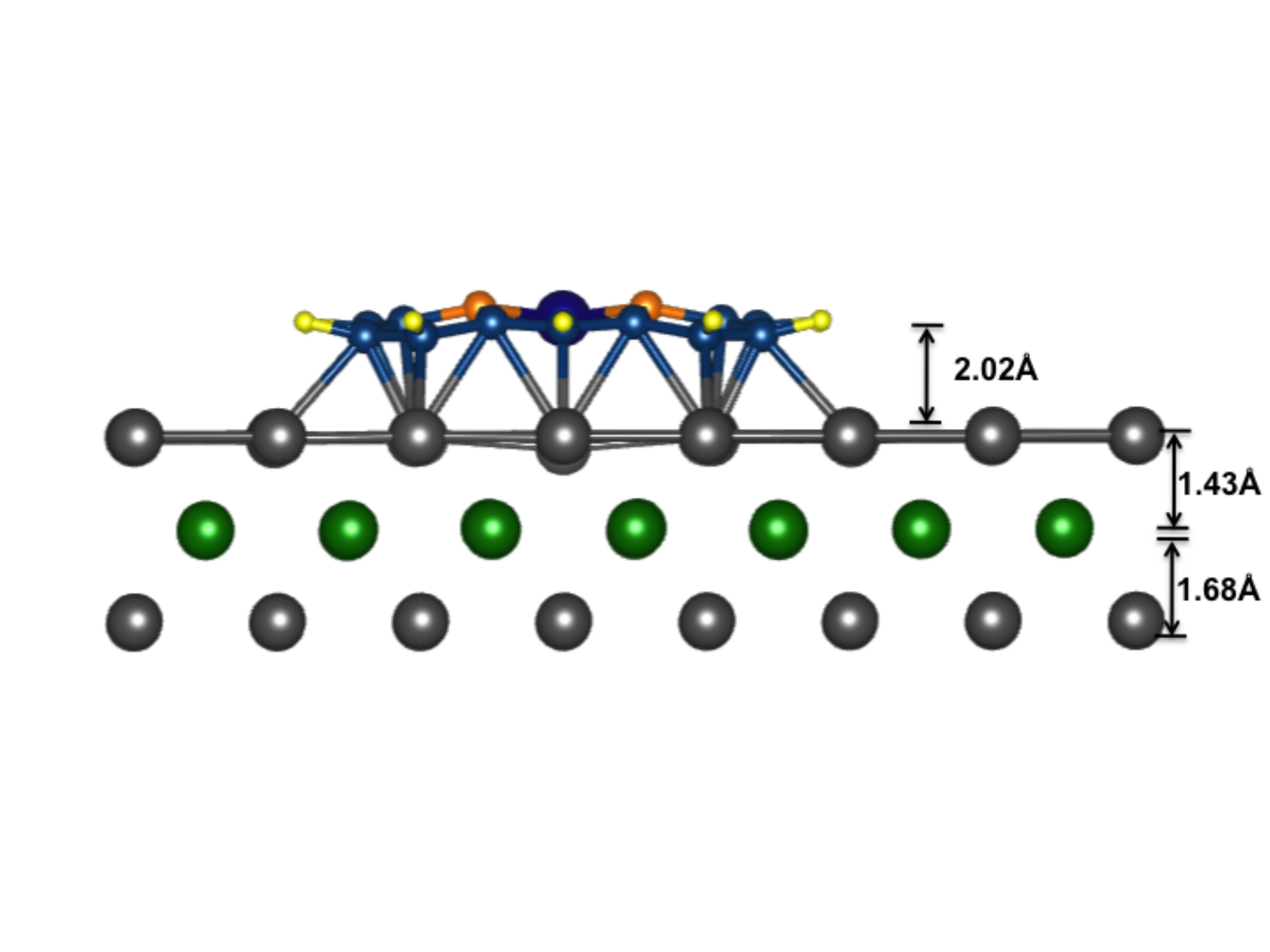}\\
 (a) \hskip 8cm (b) \\~\\
\caption{(Color online) {\sl TOP-R} configuration. (a) Top view, where Fe-N bond lengths are shown.  (b) Side view, where minimum distances between different layers are shown.} 

\label{figsi2}
\end{center}
\end{figure}

\par (ii) {\sl TOP-R}: Another possible configuration with Fe on top of a Co surface atom is denoted as a {\sl TOP-R} configuration, where FeP is rotated by 45 $^{\circ}$ around the vertical Fe-Co axis. In this situation, N's are placed at hollow sites. This prevents the formation of Co-N direct bonding and hence, the Fe-N-Co indirect exchange path becomes ineffective. The Fe-Co direct bonding distance (2.29 \AA) also changes slightly compared to that in the {\sl TOP} configuration and so does the minimum FeP-substrate distance, which becomes 2.02 \AA ~(\ref{figsi2}(b)). The resulting distortion in the Co surface layer is also different from the {\sl TOP} configuration. The Co-Co distance (2.64 \AA) right below FeP is rather stretched than being contracted. The minimum interlayer distance between the topmost and 2$^{nd}$ layer of Co is also changed to 1.43 \AA ~ while the same between 2$^{nd}$ and 3$^{rd}$ layers changes to 1.68 \AA. 

\begin{figure}[!hbp]
\begin{center}
\includegraphics[scale=0.075]{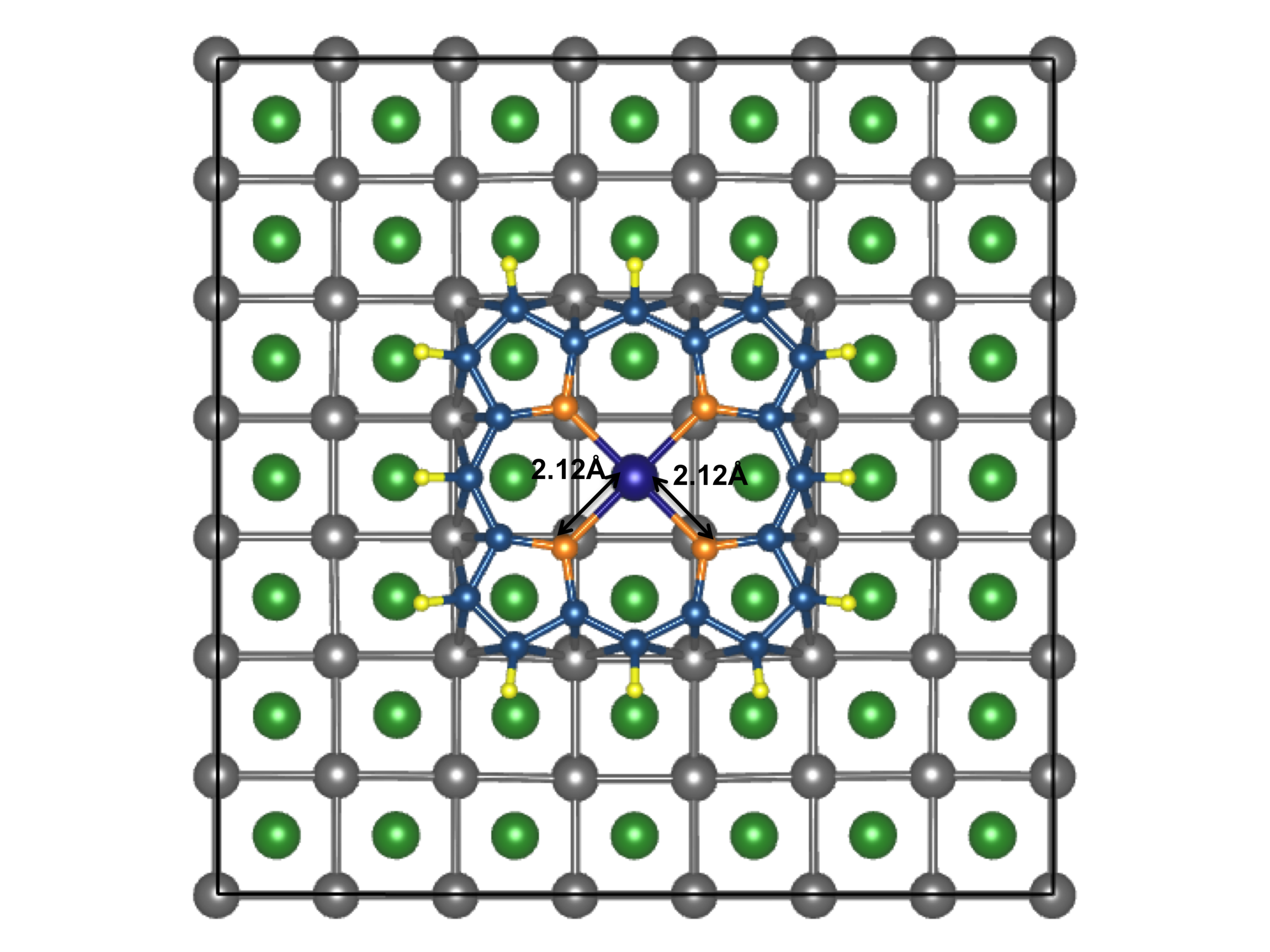} 
 \includegraphics[scale=0.075]{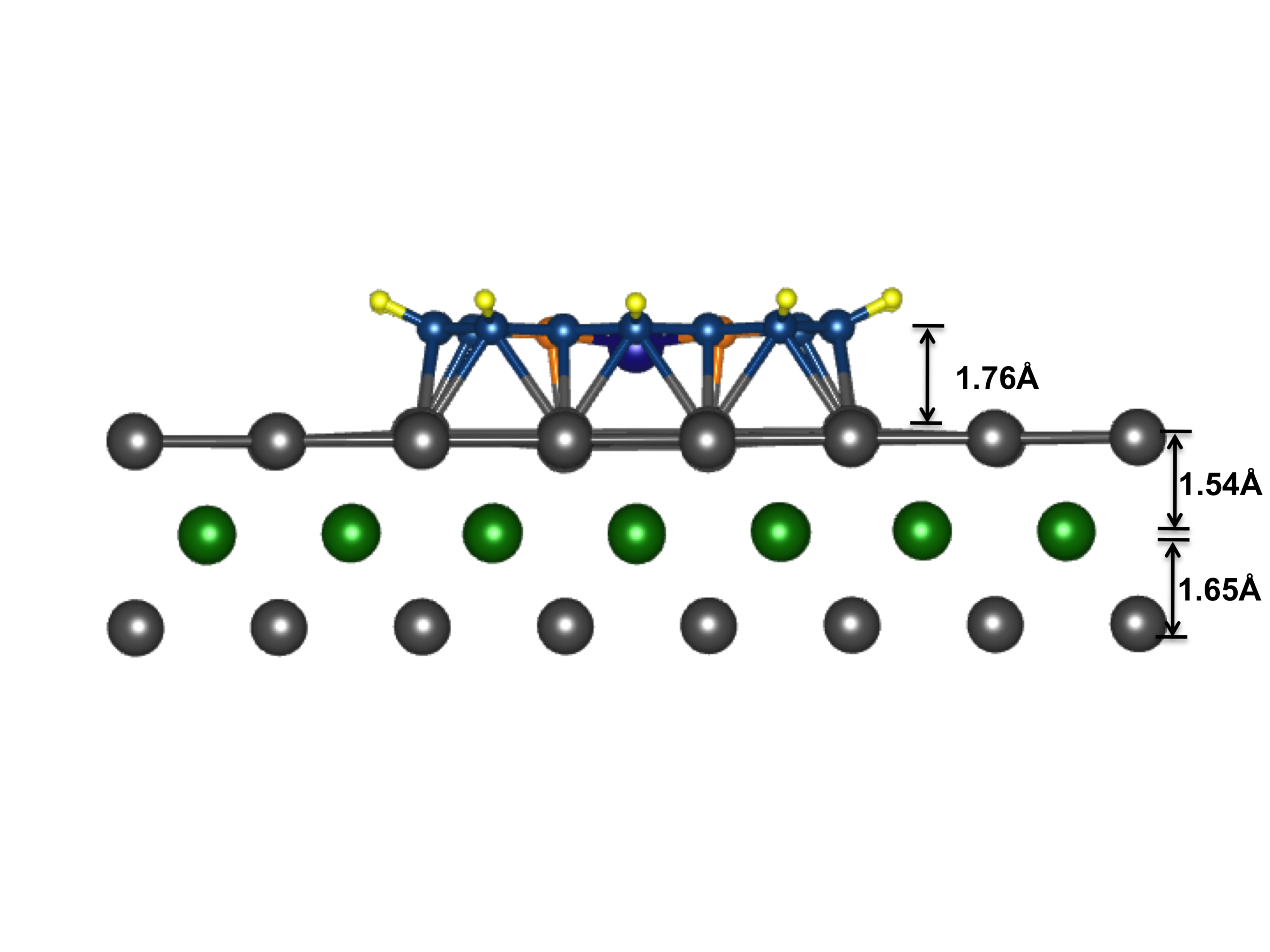}\\
 (a) \hskip 8cm (b) \\~\\
\caption{(Color online) {\sl HOLLOW} configuration. (a) Top view, where Fe-N bond lengths are shown. (b) Side view, where minimum distances between different layers are shown.}
\label{figsi3}
\end{center}
\end{figure}

\par {\sl HOLLOW}: The distortion in Co layers is considerably less in the {\sl HOLLOW} configuration depicted in \ref{figsi3}. In this case, the geometry is stabilized in such a way that the Fe-center does not sit on top of any Co atom but on a hollow site, i.e on top of a 2$^{nd}$ layer Co atom. As there is no direct bonding between Fe and Co, the surface Co layer is not much affected. The Fe-Co bond length is much larger here whereas a direct bond exists between N's and surface Co atoms. In this case, next nearest neighbor Co-Co distance (3.60 \AA) should be compared with N-Fe-N distance (4.25 \AA) (\ref{figsi3}(a)).  This forces the Fe-N-Co bond angle to be 75.35  $^{\circ}$, which is much smaller than 90  $^{\circ}$. The minimum distance between FeP and the surface layer is 1.78 \AA. The minimum distance between 1$^{st}$-2$^{nd}$ and 2$^{nd}$-3$^{rd}$ layers do not differ much (1.62 \AA and 1.66 \AA, respectively), as shown in \ref{figsi3}(b).

\begin{figure}[!hbp]
\begin{center}
\includegraphics[scale=0.075]{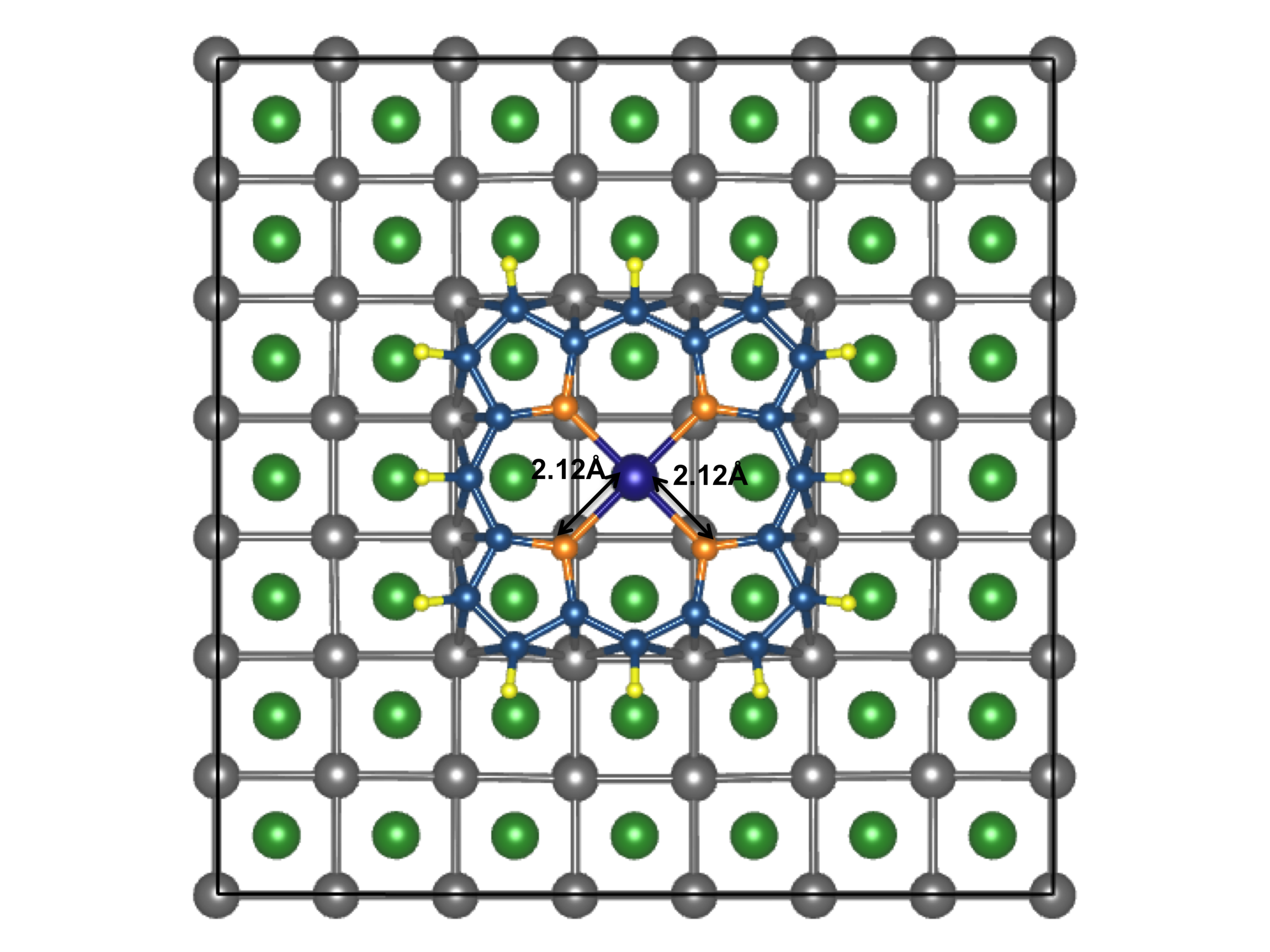} 
\includegraphics[scale=0.075]{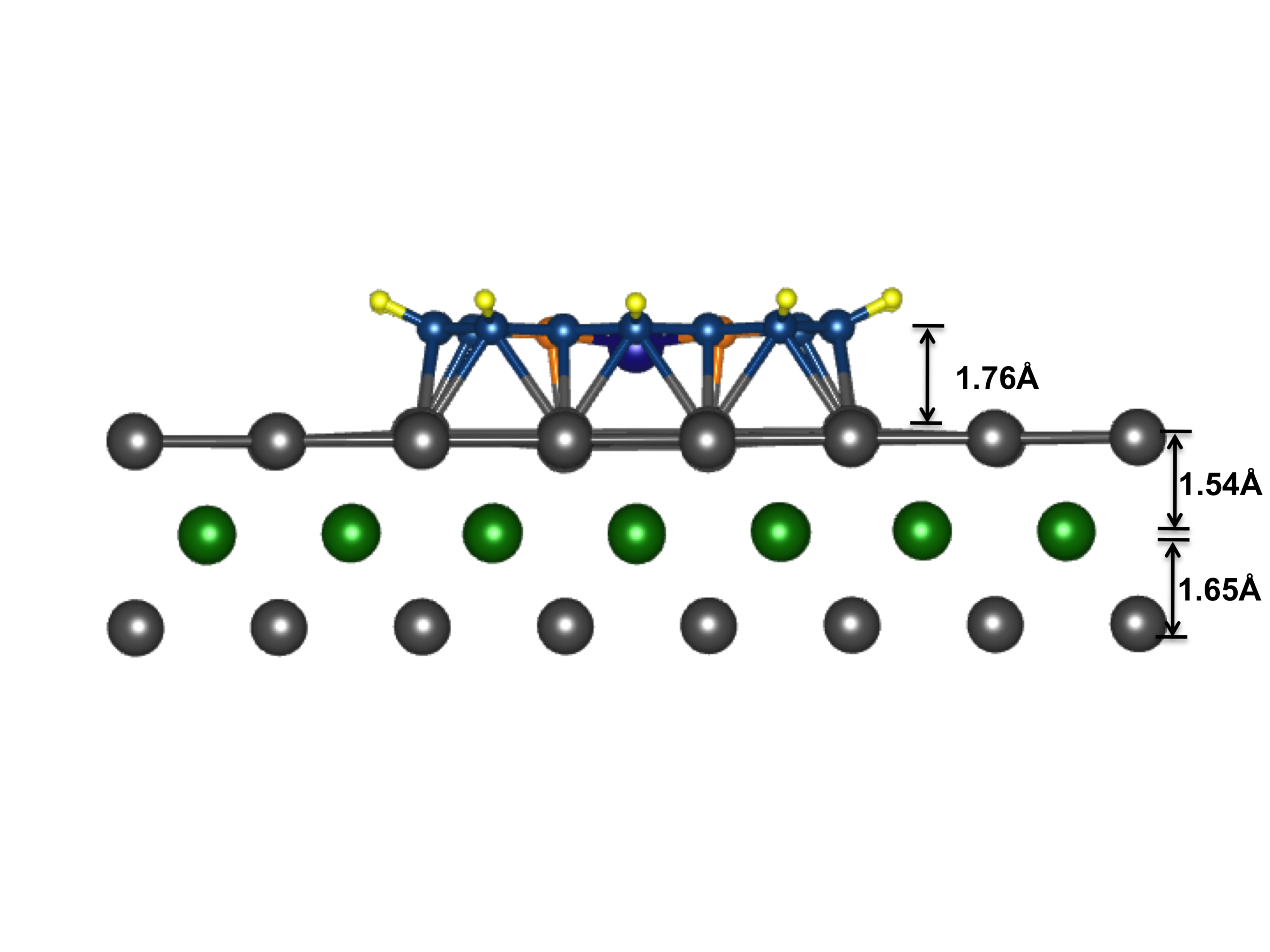}\\
 (a) \hskip 8cm (b) \\~\\
\caption{(Color online) {\sl HOLLOW-R} configuration. (a) Top view, where Fe-N bond lengths are shown. (b) Side view, where minimum distances between layers are shown.}
\label{figsi4}
\end{center}
\end{figure}

\par {\sl HOLLOW-R}: If we rotate FeP with respect to the {\sl HOLLOW} configuration by 45 $^{\circ}$ about the vertical axis (denoted as {\sl HOLLOW-R}), Co-N direct bonds will not form and as in the previous structure, Fe-Co bond is longer compared to that in TOP/TOP-R configuration. Unlike the previous one, the indirect Fe-N-Co exchange path is ineffective here.  The Fe-N bond length is  2.10 \AA ~(\ref{figsi4}(a)), which is slightly less compared to other structures but is enough to change the spin state of Fe. The minimum distance between FeP and surface Co-layer is 1.89 \AA. As the distortion is not much in the first Co layer, the minimum distances between 1$^{st}$-2$^{nd}$ and 2$^{nd}$-3$^{rd}$ layers are pretty much similar, 1.59 \AA~ and 1.66 \AA ~respectively (\ref{figsi4}(b)).

\end{document}